# Fast Set Intersection in Memory

Bolin Ding
University of Illinois at Urbana-Champaign
201 N. Goodwin Avenue
Urbana, IL 61801, USA
bding3@uiuc.edu

Arnd Christian König
Microsoft Research
One Microsoft Way
Redmond, WA 98052, USA
chrisko@microsoft.com

## ABSTRACT

Set intersection is a fundamental operation in information retrieval and database systems. This paper introduces linear space data structures to represent sets such that their intersection can be computed in a worst-case efficient way. In general, given $k$ (preprocessed) sets, with totally $n$ elements, we will show how to compute their intersection in expected time $\mathbf{O}(n/\sqrt{w} + kr)$, where $r$ is the intersection size and $w$ is the number of bits in a machine-word. In addition, we introduce a very simple version of this algorithm that has weaker asymptotic guarantees but performs even better in practice; both algorithms outperform the state of the art techniques for both synthetic and real data sets and workloads.

## 1. INTRODUCTION

Fast processing of *set intersections* is a key operation in many query processing tasks in the context of databases and information retrieval. For example, in the context of databases, set intersections are used in the context of various forms of data mining, text analytics, and evaluation of conjunctive predicates. They are also the key operations in enterprise and web search.

Many of these applications are interactive, meaning that the latency with which query results are displayed is a key concern. It has been shown in the context of search that query latency is critical to user satisfaction, with increases in latency directly leading to fewer search queries being issued and higher rates of query abandonment [10, 17]. As a consequence, significant portions of the sets to be intersected are often cached in main memory.

This paper will study the performance of set intersection algorithms for main-memory resident data. Note that these techniques are also relevant in the context of large disk-based (inverted) indexes, when large fractions of these reside in a main memory cache. There has been considerable study of set intersection algorithms in information retrieval (e.g., [12, 4, 11]). Most of these papers assume that the underlying data structure is an *inverted index* [23]. Much of this work (e.g., [12, 4]) focuses on *adaptive algorithms* which use the number of comparisons as measure of overhead. For in-memory data, additional structures which encode additional skipping-steps [18], tree-based structures [7], or hash-based algorithms become possible, which often outperform inverted indexes; e.g., using hash-based dictionaries, intersecting two sets $L_1$, $L_2$ requires expected time $\mathbf{O}(\min(|L_1|, |L_2|))$, which is a factor of $\Theta(\log(1+\max(|L_1|/|L_2|, |L_1|/|L_2|)))$ better than the best possible worst-case performance of comparison-based algorithms [6]. In this work, we propose new set intersection algorithms aimed at fast performance. These outperform the competing techniques for most inputs and are also robust in that – for inputs where they are not optimal – they are close to the best-performing algorithm. The tradeoff for this gain is a slight increase in the size of the data structures, when compared to an inverted index; however, in user-facing scenarios where latency is crucial, this tradeoff is often acceptable.

### 1.1 Contributions

Our approach leverages two key observations: (a) If $w$ is the size (in bits) of a machine-word, we can encode a set from a universe of $w$ elements in a single machine word, allowing for very fast intersections. (b) For the data distributions seen in many real-life examples (in particular search applications), the size of intersections is typically much smaller than the smallest set being intersected.

To illustrate the second observation, we analyzed the 10K most frequent queries issued against the *Bing Shopping* portal. For 94% of all queries it held that the size of the full intersection was at least one order of magnitude smaller than the document frequency of the least frequent keyword; for 76% of the queries the difference was two orders of magnitude. By exploiting these two observations, we make the following contributions.

(i) We introduce linear-space data structures to represent sets such that their intersection can be computed in a worst-case efficient way. Given $k$ sets, with $n$ elements in total, these data structures allow us to compute their intersection in expected time $\mathbf{O}(n/\sqrt{w} + kr)$, where $r$ is the size of the intersection and $w$ is the number of bits in a machine-word; when the size of the intersection is an order of magnitude (or more) smaller than the size of the smallest set being intersected, our approach yields significant improvements in execution time over previous approaches.

To the best of our knowledge, the best asymptotic bound for fast set intersection is achieved by the $\mathbf{O}\big((n(\log_2 w)^2)/w + kr\big)$ algorithm of [6]. However, note that the bound relies on a large value of $w$; in practice, $w$ is small (and constant), and $w < 2^{16} = 65536$ bits implies $1/\sqrt{w} < (\log_2 w)^2/w$. More importantly, [6] requires complex bit-manipulation, making it slow in practice, which we will demonstrate empirically in Section 4.

(ii) We describe a much simpler algorithm that computes the intersection in expected $\mathbf{O}(n/\alpha^m + mn/\sqrt{w} + kr\sqrt{w})$ time, where $\alpha$ is a constant determined by $w$, and $m$ is a parameter. This algorithm has weaker guarantees in theory, but performs better in practice, and gives significant improvements over the various data structures typically used, while being very simple to implement.

## 2. BACKGROUND AND RELATED WORK

**Algorithms based on Ordered Lists:** Most work on set intersection focuses on ordered lists as the underlying data structure, in particular algorithms using *inverted indexes*, which have become the standard data structure in information retrieval. Here, documents are identified via a *document ID*, and for each term $t$, the inverted index stores a sorted list of all document IDs containing $t$.

Using this representation, two sets $L_1, L_2$ of similar sizes (i.e., $|L_1| \approx |L_2|$) can be intersected efficiently using a linear merge by scanning both lists in parallel, requiring $\mathbf{O}(|L_1| + |L_2|)$ operations (the "merge step" in merge sort). This approach is wasteful when set sizes differ significantly or only small fractions of the sets intersect. For very different set sizes, algorithms have been proposed that exploit this asymmetry, requiring $\log \binom{|L_1|+|L_2|}{|L_1|} + |L_1|$ comparisons at most (for $|L_1| < |L_2|$) [16].

To improve the performance further, there has recently been significant work on so-called *adaptive* set-intersection algorithms for set intersections [12, 4, 13, 1, 2, 5]. These algorithms use the total number of comparisons as measure of the algorithm's complexity and aim to use a number of comparisons as close as possible to the minimum number of comparisons ideally required to establish the intersection. However, the resulting reduction in the number of comparisons does not necessarily result in performance improvements in practice: for example, in [2], binary search based algorithms outperform a parallel scan only when $|L_2| < 20|L_1|$, even though several times fewer comparisons are needed.

**Hierarchical Representations:** There are various algorithms for set intersections based on variants of balanced trees (e.g. [9], treaps [7], and skip-lists [18]), computing the intersection of (preprocessed) sets $L_1, L_2$ in $\mathbf{O}(|L_1| \log(|L_2|/|L_1|))$ (for $|L_1| < |L_2|$) operations. However, while some form of "skipping" is commonly used as part of algorithms based on inverted indexes, skip-lists (or trees) are typically not used in the scenarios outlined above (with static set data) due to the required space-overhead. A novel and compact two-level representation of posting lists aimed at fast intersections in main memory was proposed in [19].

**Algorithms based on Hashing:** Using a hash-based representation of sets can speed up the intersection of sets $L_1, L_2$ with $|L_1| \ll |L_2|$ significantly (expected time $\mathbf{O}(|L_1|)$ – by looking up all elements of $L_1$ in the hash-table of $L_2$); however, because of the added indirection, this approach performs poorly for less skewed set sizes. A new hashing-based approach is proposed in [6]: here, the elements in sets $L_1, L_2$ are mapped using a hash-function $h$ to smaller (approximate) representations $h(L_1), h(L_2)$. These representations are then intersected to compute $H = h(L_1) \cap h(L_2)$. Finally, the set of all elements in the original sets that map to $H$ via $h$ are computed and any "false positives" removed. As the hashed images $h(L_1), h(L_2)$ to be intersected are smaller than the original sets (using fewer bits), they can be intersected more quickly. Given $k$ sets of total size $n$, their intersection can be computed in expected time $\mathbf{O}\big((n \log^2 w)/w + kr\big)$, where $r = |\bigcap_i L_i|$.

**Score-based pruning:** In many IR engines it is possible to avoid computing full intersections by leveraging scoring functions that are monotonic in the individual term-wise scores; this makes it possible to terminate the intersection processing early using approaches such as TA [15] or *document-at-a-time* (DAAT) processing (e.g., [8]). However, in practice, this is often not possible, either because of the complexity of the scoring function (e.g., non-monotonic machine-learning based ranking functions) or because full intersection results are required. Our approach is based on partitioning the elements in each set into very small ($\approx 8$ elements) groups, for which we have fast intersection schemes. Hence, DAAT-approaches can be combined with our work by using these small groups in place of individual documents.

**Set intersections using multiple cores:** Techniques that exploit multi-core architectures to speed up set intersections are described in [20, 22]. The use of multiple cores is orthogonal to our approach in the sense that our algorithms can be parallelized for these architectures as well; however, this is beyond the scope of our paper.

## 3. OUR APPROACH

**Notation:** We are given a collection of $N$ sets $\mathcal{S} = \{L_1, \ldots, L_N\}$, where $L_i \subseteq \Sigma$ and $\Sigma$ is the universe of elements in the sets; let $n_i = |L_i|$ be the size of set $L_i$. Suppose elements in a set are ordered, and for a set $L$, let $\inf(L)$ and $\sup(L)$ be the minimum and maximum elements of a set $L$, respectively. We use $w$ to denote the size (number of bits) of a word on the target processor. Throughout the paper we will use $\log$ to denote $\log_2$. Finally, we use $[w]$ to denote the set $\{1, \ldots, w\}$. Our approach can be extended to bag semantics by additionally storing element frequency.

**Framework:** Our task is to design data structures such that the intersection of multiple sets can be computed efficiently. We differentiate between a *pre-processing stage*, during which we reorganize each set and attach additional index structures, and an *online processing stage*, which uses the pre-processed data structures to compute intersections. An intersection query is specified via a collection of $k$ sets $L_1, L_2, \ldots, L_k$ (to simplify notations, we use the offsets $1, 2, \ldots, k$ to refer to the sets in a query throughout this section); our goal is to compute $L_1 \cap L_2 \cap \ldots \cap L_k$ efficiently. Note that pre-processing is typical of most non-trivial data structures used for computing set intersections; even building simple non-compressed inverted indexes requires sorting the posting lists as a pre-processing step. We require the pre-processing stage to be time/space-efficient in that it does not require more than $\mathbf{O}(n_i \log n_i)$ time (necessary for sorting) and linear space $\mathbf{O}(n_i)$.

The size of intersection $|L_1 \cap L_2|$ is a lower bound of the time needed to compute the intersection. Our method leverages two key ideas to approach this lower bound: (i) The intersection of two sets in a *small universe* can be computed very efficiently; in particular, if the two sets are subsets of $\{1, 2, \ldots, w\}$, we can encode them as single machine-words and compute their intersection using a bitwise-AND. (ii) A small number of elements in a large universe can be mapped into a small universe.

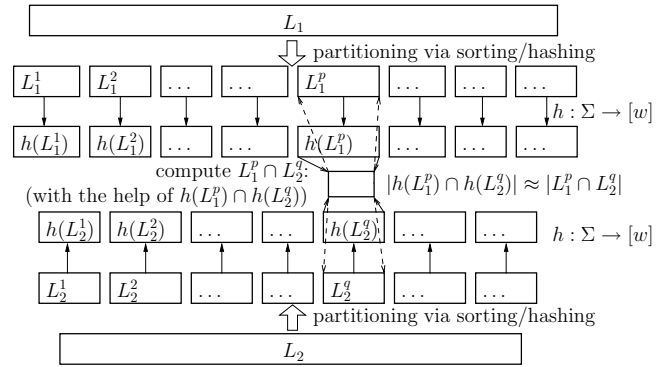

**Figure 1: Algorithmic Framework**

We leverage these two ideas by first partitioning each set $L_i$ into smaller *groups* $L_i^j$'s, which are intersected separately. In the pre-processing stage, we map each small group into a small universe $[w] = \{1, 2, \ldots, w\}$ using a universal hash function $h$ and encode the image $h(L_i^j)$ with a machine-word. Then, in the on-line processing stage, to compute the intersection of two small groups $L_1^p$ and $L_2^q$, we first use a bitwise-AND operation to compute $H = h(L_1^p) \cap h(L_2^q)$, and then try to "recover" $L_1^p \cap L_2^q$ using the inverse mapping $h^{-1}$ from $H$. The union of $L_1^p \cap L_2^q$'s forms $L_1 \cap L_2$. Moreover, if the intersection $L_1 \cap L_2$ is of a small size compared to $|L_1|$ and $|L_2|$ (seen in practice), a large fraction of the small groups with overlapping ranges has an empty intersection; thus, by using the word-representations of $H$ to detect these groups quickly, we can skip much unnecessary computation, resulting in significant speed-up. The resulting algorithmic framework is illustrated in Figure 1. Given this overall approach, the key questions become how to form groups, what structures to be used to represent them, and how to process intersections of these small groups.

We will discuss these details in the following sections. All the formal proofs of analytical results are deferred to the appendix.

### 3.1 Intersection via Fixed-Width Partitions

We first consider the case when there are only two sets $L_1$ and $L_2$ in the intersection query. We will present a pair of pre-processing and online processing algorithms, which we use to illustrate the basic ideas of our algorithms. We subsequently refine and extend our techniques to $k$ sets in Section 3.2.

In the *pre-processing stage*, $L_1$ and $L_2$ are sorted, and partitioned into groups (recall $w$ is the word width)

$$L_1^1, L_1^2, \ldots, L_1^{\lceil n_1/\sqrt{w} \rceil}, \text{ and } L_2^1, L_2^2, \ldots, L_2^{\lceil n_2/\sqrt{w} \rceil}$$

of equal size $\sqrt{w}$ (except the last ones).

In the *online processing stage* (Algorithm 1), the small groups are scanned in order. If the ranges of $L_1^p$ and $L_2^q$ overlap, we may have $L_1^p \cap L_2^q \neq \emptyset$. The intersection $L_1^p \cap L_2^q$ of each pair of overlapping groups is computed (line 8) in some iteration. And finally, the union of all these intersections is $L_1 \cap L_2$. Since each group is scanned once, line 2-10 repeat for $\mathbf{O}((n_1 + n_2)/\sqrt{w})$ iterations.

The major remaining question now becomes how to compute $L_1^p \cap L_2^q$ efficiently with proper pre-processing? For this purpose, we map each group $L_1^p$ or $L_2^q$ into a small universe for fast intersection, and we leverage *single-word representations* to store and manipulate sets from a small universe.

**Single-Word Representation of Sets:** We represent a set $A \subseteq [w] = \{1, 2, \ldots, w\}$ using a single machine-word of width $w$ by setting the $y$-th bit as 1 iff $y \in A$. We refer to this as the *word representation* $w(A)$ of $A$. For two sets $A$ and $B$, the bitwise-AND $w(A) \wedge w(B)$ (computed in $\mathbf{O}(1)$ time) is the word representation of $A \cap B$. Given a word representation $w(A)$, all the elements of $A$ can be retrieved in linear time $\mathbf{O}(|A|)$ [1]. In the rest of this paper, if $A \subseteq [w]$, we use $A$ to denote both a set and its word representation.

**Pre-processing Stage:** Elements in a set $L_i$ are sorted as $\{x_i^1, x_i^2, \ldots, x_i^{n_i}\}$ (i.e., $x_i^k < x_i^{k+1}$) and $L_i$ is partitioned as follows:

$$L_i^1 = \{x_i^1, \ldots, x_i^{\sqrt{w}}\}, \quad L_i^2 = \{x_i^{\sqrt{w}+1}, \ldots, x_i^{2\sqrt{w}}\}, \ldots \quad (1)$$

$$L_i^j = \{x_i^{(j-1)\sqrt{w}+1}, x_i^{(j-1)\sqrt{w}+2}, \ldots, x_i^{j\sqrt{w}}\}, \ldots \quad (2)$$

For each small group $L_i^j$, we compute the word-representation of its image under a universal hash function $h : \Sigma \to [w]$, i.e., $h(L_i^j) = \{h(x) \mid x \in L_i^j\}$. In addition, for each position $y \in [w]$ and each small group $L_i^j$, we also maintain the *inverted mapping* $h^{-1}(y, L_i^j) = \{x \mid x \in L_i^j \text{ and } h(x) = y\}$, i.e., for each $y \in [w]$ we store the elements in $L_i^j$ with hash value $y$, in a short list which supports ordered access. We ensure that the order of these elements is identical across different $h^{-1}(y, L_i^j)$'s and $L_i$'s; in this way, we can intersect these short lists using a linear merge.

---

[1] We use the following well-known technique: ($\oplus$ is bitwise-XOR)
(i) lowbit $= ((w(A) - 1) \oplus w(A)) \wedge w(A)$ is the lowest 1-bit of $w(A)$. For the smallest element $y$ in $A$, we have $2^y = $ lowbit.
$y = \log(\text{lowbit}) \in A$ can be computed using the machine instruction NLZ (number of leading zeros) or pre-computed lookup tables.
(ii) Set $w(A)$ as $w(A) \oplus$ lowbit and repeat (i) to scan the next smallest element until $w(A)$ becomes 0.

EXAMPLE 3.1. (PRE-PROCESSING AND DATA STRUCTURES)
*Suppose we have two sets $L_1 = \{1001, 1002, 1004, 1009, 1016, 1027, 1043\}$, $L_2 = \{1001, 1003, 1005, 1009, 1011, 1016, 1022, 1032, 1034, 1049\}$. And, let $w = 16$ ($\sqrt{w} = 4$). For simplicity, $h$ is selected to be $h(x) = (x - 1000) \mod 16$. $L_1$ is partitioned into 2 groups: $L_1^1 = \{1001, 1002, 1004, 1009\}$, $L_1^2 = \{1016, 1027, 1043\}$, and $L_2$ is partitioned into 3 groups: $L_2^1 = \{1001, 1003, 1005, 1009\}$, $L_2^2 = \{1011, 1016, 1022, 1032\}$, $L_2^3 = \{1034, 1049\}$. We pre-compute: $h(L_1^1) = \{1, 2, 4, 9\}$, $h(L_1^2) = \{0, 11\}$, $h(L_2^1) = \{1, 3, 5, 9\}$, $h(L_2^2) = \{0, 6, 11\}$, $h(L_2^3) = \{1, 2\}$. We also pre-process $h^{-1}(y, L_i^p)$'s: for example, $h^{-1}(0, L_1^2) = \{1016\}$, $h^{-1}(0, L_2^2) = \{1016, 1032\}$, $h^{-1}(11, L_1^2) = \{1027, 1043\}$, and $h^{-1}(11, L_2^2) = \{1011\}$.* □

---

1: $p \leftarrow 1, q \leftarrow 1, \Delta \leftarrow \emptyset$
2: **while** $p \leq n_1$ and $q \leq n_2$ **do**
3:   **if** $\inf(L_2^q) > \sup(L_1^p)$ **then**
4:     $p \leftarrow p + 1$
5:   **else if** $\inf(L_1^p) > \sup(L_2^q)$ **then**
6:     $q \leftarrow q + 1$
7:   **else**
8:     compute $(L_1^p \cap L_2^q)$ using IntersectSmall
9:     $\Delta \leftarrow \Delta \cup (L_1^p \cap L_2^q)$
10:     **if** $\sup(L_1^p) < \sup(L_2^q)$ **then** $p \leftarrow p + 1$ **else** $q \leftarrow q + 1$
11: $\Delta$ is the result of $L_1 \cap L_2$

**Algorithm 1:** Intersection via fixed-width partitioning

**Online Processing Stage:** The algorithm used to intersect two sets is shown in Algorithm 1. Since elements in $L_i$ are sorted, Algorithm 1 ensures that if the ranges of any two small groups $L_1^p, L_2^q$ overlap, their intersection is computed (line 8). After scanning all such pairs, $\Delta$ must then contain the intersection of the whole sets.

Now the question is: how to compute the intersection of two small groups $L_1^p \cap L_2^q$ efficiently? For this purpose, we introduce the algorithm IntersectSmall (Algorithm 2), which:
(i) first computes $H = h(L_1^p) \cap h(L_2^q)$ using a bitwise-AND;
(ii) for each (1-bit) $y \in H$, intersects the corresponding inverted mappings using the linear merge algorithm.

---

IntersectSmall($L_1^p, L_2^q$): computing $L_1^p \cap L_2^q$
1: Compute $H \leftarrow h(L_1^p) \cap h(L_2^q)$
2: **for** each $y \in H$ **do**
3:   $\Gamma \to \Gamma \cup (h^{-1}(y, L_1^p) \cap h^{-1}(y, L_2^q))$
4: $\Gamma$ is the result of $L_1^p \cap L_2^q$

**Algorithm 2:** Computing the intersection of small groups

EXAMPLE 3.2. (ONLINE PROCESSING)
*Following Example 3.1, to compute $L_1 \cap L_2$, we need to compute $L_1^1 \cap L_2^1$, $L_1^2 \cap L_2^2$, and $L_1^2 \cap L_2^3$ (pairs with overlapping ranges): for example, for computing $L_1^2 \cap L_2^2$, we first compute $h(L_1^2) \cap h(L_2^2) = \{0, 11\}$; then $L_1^2 \cap L_2^2 = \bigcup_{y=0,11} (h^{-1}(y, L_1^2) \cap h^{-1}(y, L_2^2)) = \{1016\}$. Similarly, we can compute $L_1^1 \cap L_2^1 = \{1001, 1009\}$. Finally, we find $h(L_1^2) \cap h(L_2^3) = \emptyset$, and thus $L_1^2 \cap L_2^3 = \emptyset$. So, we have $L_1 \cap L_2 = \{1001, 1009\} \cup \{1016\} \cup \emptyset$.* □

Note that word representations and inverted mappings for $L_i$ are pre-computed, and word-representations can be intersected using one operation. So the running time of IntersectSmall is bounded by the number of pairs of elements, one from $L_1^p$ and one from $L_2^q$, that are mapped to the same hash-value. This number can be shown to be equal (in expectation) to the intersection size plus $\mathbf{O}(1)$ for each group $L_i^j$. Using this, we obtain Algorithm 1's running time:

THEOREM 3.3. *Algorithm 1 computes $L_1 \cap L_2$ in expected* $\mathbf{O}\left(\frac{n_1+n_2}{\sqrt{w}} + r\right)$ *time, where $r = |L_1 \cap L_2|$.*

To achieve a better bound, we optimize the group sizes: with $L_1$ and $L_2$ partitioned into groups of sizes $s_1^* = \sqrt{wn_1/n_2}$ and $s_2^* = \sqrt{wn_2/n_1}$, respectively, $L_1 \cap L_2$ can be computed in expected $\mathbf{O}(\sqrt{n_1 n_2/w} + r)$ time. A detailed analysis of the effect of group size on running times can be found in Section A.1.1.

**Overhead of Pre-processing:** If only the bound in Theorem 3.3 is required, then to pre-process a set $L_i$ of size $n_i$, it is obvious that $\mathbf{O}(n_i \log n_i)$ time and $\mathbf{O}(n_i)$ space suffice: we only need to partition a sorted list into small groups of size $\sqrt{w}$, and for each small group, construct the word representation and inverted mapping in linear time using the hash function $h$.

To achieve the better bound $\mathbf{O}(\sqrt{n_1 n_2/w}+r)$, we need multiple "resolutions" of the partitioning of a set $L_i$. This is because, as discussed above, the optimal group size $s_1^* = \sqrt{wn_1/n_2}$ of the set $L_1$ also depends on the size $n_2$ of the set $L_2$ to be intersected with it. For this purpose, we partition a set $L_i$ into small groups of size $2, 4, \ldots, 2^j$, etc. To compute $L_1 \cap L_2$ for the given two sets, suppose $s_i^*$ is the optimal group size of $L_i$; we then select the actual group size $s_i^{**} = 2^t$ s.t. $s_i^* \leq s_i^{**} \leq 2s_i^*$, obtaining the same bound. A carefully-designed multi-resolution data structure enabling access to these groups consumes only $\mathbf{O}(n_i)$ space for $L_i$. We will describe and analyze this structure in Section 3.2.1.

THEOREM 3.4. *To pre-process a set $L_i$ of size $n_i$ for Algorithm 1, we need $\mathbf{O}(n_i \log n_i)$ time and $\mathbf{O}(n_i)$ space (in words).*

**Limitations of Fixed-Width Partitions:** The main limitation of the proposed approach is that it is difficult to extend to more than two sets, because the partitioning scheme we use is not well-aligned for more than two sets: for three sets, e.g., there may be more than $\mathbf{O}((n_1 + n_2 + n_3)/\sqrt{w})$ triples of small groups that overlap. We introduce a different partitioning scheme to address this issue in Section 3.2, which extends to $k > 2$ sets.

## 3.2 Intersection via Randomized Partitions

In this section, we will introduce an algorithm based on a randomized partitioning scheme to compute the intersection of two or more sets. The general approach is as follows: instead of fixed-width partitions, we use a hash function $g$ to partition each set into small groups, using the most significant bits of $g(x)$ to group an element $x \in \Sigma$. This reduces the number of combinations (pairs) of small groups we have to intersect, allowing us to prove bounds similar to Theorem 3.3 for computing intersections of $k > 2$ sets.

**Pre-processing Stage:** Let $g$ be a hash function $g : \Sigma \to \{0, 1\}^w$ mapping an element to a bit-string (or binary number); we use $g_t(x)$ to denote the $t$ most significant bits of $g(x)$. We say that for two bit-strings $z_1$ and $z_2$, $z_1$ is a $t_1$-prefix of $z_2$, iff $z_1$ is identical to the highest $t_1$ bits in $z_2$; e.g., 1010 is a 4-prefix of 101011.

To pre-process a set $L_i$, we partition it into groups $L_i^z = \{x \mid x \in L_i \text{ and } g_t(x) = z\}$ for all $z \in \{0, 1\}^t$ (some $t$). As before, we compute the word representation of the image of each $L_i^z$ under another hash function $h : \Sigma \to [w]$, and inverted mappings $h^{-1}$.

**Online Processing Stage:** This stage is similar to our previous algorithm: to compute the intersection of two sets $L_1$ and $L_2$, we compute the intersections of *pairs of overlapping small groups*, one from each set, and finally take the union of these intersections.

In general, suppose $L_1$ is partitioned using $g_{t_1} : \Sigma \to \{0, 1\}^{t_1}$, and $L_2$ is partitioned using $g_{t_2} : \Sigma \to \{0, 1\}^{t_2}$. Assume $n_1 \leq n_2$ and $t_1 \leq t_2$. We now intersect sets $L_1$ and $L_2$ using Algorithm 3. The major improvement of Algorithm 3 compared to Algorithm 1 is that in Algorithm 1, we need compute $L_1^p \cap L_2^q$ when the ranges of $L_1^p$ and $L_2^q$ overlap; in Algorithm 3, we compute $L_1^{z_1} \cap L_2^{z_2}$ (also using Algorithm 2) *when $z_1$ is a $t_1$-prefix of $z_2$* (this is a necessary condition for $L_1^{z_1} \cap L_2^{z_2} \neq \emptyset$; so Algorithm 3 is correct). This significantly reduces the number of pairs to be intersected.

1: **for** each $z_2 \in \{0, 1\}^{t_2}$ **do**
2:    Let $z_1 \in \{0, 1\}^{t_1}$ be the $t_1$-prefix of $z_2$
3:    Compute $L_1^{z_1} \cap L_2^{z_2}$ using IntersectSmall($L_1^{z_1}, L_2^{z_2}$)
4:    Let $\Delta \leftarrow \Delta \cup (L_1^{z_1} \cap L_2^{z_2})$
5: $\Delta$ is the result of $L_1 \cap L_2$

**Algorithm 3:** 2-list Intersection via Randomized Partitioning

Based on the choices of parameters $t_1$ and $t_2$, we can either partition $L_1$ and $L_2$ into the same number of small groups (yielding the bound of Theorem 3.5), or into small groups of the (approximately) identical sizes (yielding Theorem 3.6).

THEOREM 3.5. *Algorithm 3 computes $L_1 \cap L_2$ in expected* $\mathbf{O}\left(\frac{\sqrt{n_1 n_2}}{\sqrt{w}} + r\right)$ *time ($r = |L_1 \cap L_2|$), with $t_1 = t_2 = \lceil \log \sqrt{\frac{n_1 n_2}{w}} \rceil$.*

THEOREM 3.6. *Algorithm 3 computes $L_1 \cap L_2$ in expected* $\mathbf{O}\left(\frac{n_1+n_2}{\sqrt{w}} + r\right)$ *time ($r = |L_1 \cap L_2|$), using $t_1 = \lceil \log(n_1/\sqrt{w}) \rceil$ and $t_2 = \lceil \log(n_2/\sqrt{w}) \rceil$.*

Note that when $n_1 \neq n_2$, Theorem 3.5 has a better bound than Theorem 3.6. But we can extend Theorem 3.6 to $k$-set intersection.

**Extension to More Than Two Sets:** Suppose we want to compute the intersection of $k$ sets $L_1, \ldots, L_k$, where $n_i = |L_i|$ and $n_1 \leq n_2 \leq \ldots \leq n_k$. $L_i$ is partitioned into groups $L_i^z$'s using $g_{t_i} : \Sigma \to \{0, 1\}^{t_i}$. Note that $g_{t_i}$'s are generated from the same hash function $g$. We use $t_i = \lceil \log(n_i/\sqrt{w}) \rceil$ and proceed as in Algorithm 4.

Algorithm 4 is almost identical to Algorithm 3, but is generalized to $k$ sets: for each $z_k \in \{0, 1\}^{t_k}$, we pick the group identifiers $z_i$ to be the $t_i$-prefix of $z_k$, and we only intersect groups $L_1^{z_1}, L_2^{z_2}, \ldots, L_k^{z_k}$, where $z_1, z_2, \ldots, z_k$ share a prefix of size $t_1$. Also, we extend IntersectSmall (Algorithm 2) for $k$ groups: we first compute the intersection (bitwise-AND) of hash images (their word-representations) of the $k$ groups $L_i^{z_i}$'s; and, if the result $H = \bigcap_{i=1}^{k} h(L_i^{z_i})$ is not zero, for each (1-bit) $y \in H$, we intersect the $k$ corresponding inverted mappings $h^{-1}(y, L_i^{z_i})$'s. Details and analysis are deferred to the appendix.

THEOREM 3.7. *Using $t_i = \lceil \log(n_i/\sqrt{w}) \rceil$, Algorithm 4 computes the intersection $\bigcap_{i=1}^{k} L_i$ of $k$ sets in expected $\mathbf{O}(n/\sqrt{w} + kr)$ time, where $r = \left|\bigcap_{i=1}^{k} L_i\right|$ and $n = \sum_{i=1}^{k} n_i = \sum_{i=1}^{k} |L_i|$.*

1: **for** each $z_k \in \{0, 1\}^{t_k}$ ($t_i = \lceil \log(n_i/\sqrt{w}) \rceil$) **do**
2:    Let $z_i$ be the $t_i$-prefix of $z_k$ for $i = 1, \ldots, k-1$
3:    Compute $\bigcap_{i=1}^{k} L_i^{z_i}$ using extended IntersectSmall
4:    Let $\Delta \leftarrow \Delta \cup (\bigcap_{i=1}^{k} L_i^{z_i})$
5: $\Delta$ is the result of $\bigcap_{i=1}^{k} L_i$

**Algorithm 4:** $k$-list Intersection via Randomized Partitioning

### 3.2.1 A Multi-resolution Data Structure

Recall that in some algorithms (e.g., Theorem 3.5), the selection of the number of small groups used for a set $L_i$ depends on the (size of) other sets being intersected with $L_i$. So by naively pre-computing the required structures for each possible group size, we would incur excessive space requirements. In this section, we describe a data structure that supports access to partitions of $L_i$ into $2^t$ groups for any possible $t$, using only $\mathbf{O}(n_i)$ space. It is illustrated in Figure 2. To support the algorithms introduced so far, this structure must also allow us:
(i) for each $L_i^z$, to retrieve the *word-representation* $h(L_i^z)$, and

258

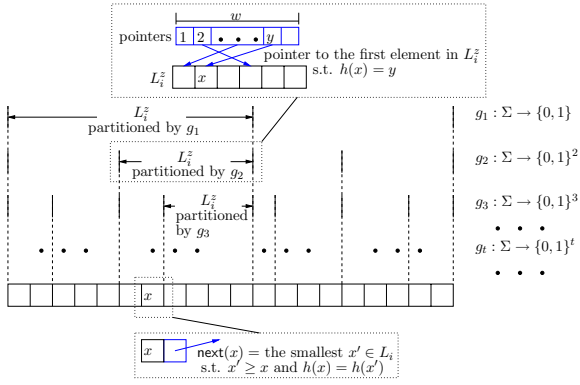

**Figure 2:** Multi-Resolution Partition of $L_i$

(ii) for each $y \in [w]$, to access all elements in $h^{-1}(y, L_i^z) = \{x \mid x \in L_i^z \text{ and } h(x) = y\}$ in time linear in its size $|h^{-1}(y, L_i^z)|$.

**Multi-resolution Partitioning:** For the ease of explanation, we suppose $\Sigma = \{0,1\}^w$ and choose $g$ as a random permutation of $\Sigma$. To pre-process $L_i$, we first order all the elements $x \in L_i$ according to $g(x)$. Then any small group $L_i^z = \{x \mid x \in L_i \text{ and } g_t(x) = z\}$ forms a consecutive interval in $L_i$ (partitions of different resolutions are formed for $t = 1, 2, \ldots$).

Note: in all of our algorithms, *universal hash functions* and *random permutations* are almost interchangeable (when used as $g$) – the differences being that (i) a permutation induces a total ordering of elements (in this data structure, this property is required), whereas hashing may result in collisions (which we can overcome by using the pre-image to break ties) and (ii) there is a slight difference in the resulting probability of, e.g., elements being grouped together (hashing results in (limited) independence, whereas permutations result in negative dependence – we account for this by using the weaker condition in our proofs).

**Word Representations of Hash Mappings:** Now, for each small group $L_i^z$, we need to pre-compute and store the word representation $h(L_i^z)$. Note the total number of small groups is $n_i/2 + n_i/4 + \ldots + n_i/2^t + \ldots \leq n_i$. So this requires $\mathbf{O}(n_i)$ space.

**Inverted Mappings:** We need to access all elements in $h^{-1}(y, L_i^z)$ in order, for each $y \in [w]$. If we were to store these mappings for each $L_i^z$ explicitly, this would require $\mathbf{O}(n_i \log n_i)$ space. However, by storing the inverted mappings $h^{-1}(y, L_i^z)$'s implicitly, we can do better, as follows:

For each group $L_i^z$, since it corresponds to an interval in $L_i$, we can store the starting and ending positions in $L_i$, denoted by $\mathsf{left}(L_i^z)$ and $\mathsf{right}(L_i^z)$. These allow us to determine if an element $x$ belongs to $L_i^z$. Now, to enable the ordered access to the inverted mappings, we define, for each $x \in L_i$, $\mathsf{next}(x)$ to be the "next" element $x'$ to $x$ on the right s.t. $h(x') = h(x)$ (i.e., with minimum $g(x') > g(x)$ s.t. $h(x') = h(x)$). Then, for each $L_i^z$ and each $y \in [w]$, we store the position $\mathsf{first}(y, L_i^z)$ of the first element $x''$ in $L_i^z$ s.t. $h(x'') = y$. Now, to access all elements in $h^{-1}(y, L_i^z)$ in order, we can start from the element at $\mathsf{first}(y, L_i^z)$, and follow the pointers $\mathsf{next}(x)$, until passing the right boundary $\mathsf{right}(L_i^z)$. And, in this way, all elements in the inverted mapping are retrieved in the same order as $g(x)$ – which we require for IntersectSmall.

**Space Requirements:** For all groups of different sizes, the total space for storing $h(L_i^z)$'s, $\mathsf{left}(L_i^z)$'s, $\mathsf{right}(L_i^z)$'s, $\mathsf{first}(y, L_i^z)$'s and $\mathsf{next}(x)$'s is $\mathbf{O}(n_i)$. So the whole multi-resolution data structure requires $\mathbf{O}(n_i)$ space. A detailed analysis is in the appendix.

When the group size $t_i$ depends only on $n_i$ (e.g., in Algorithm 4), "single-resolution" in pre-processing suffices, and the above multi-resolution scheme (for selecting $t_i$ online) is not necessary.

THEOREM 3.8. *To pre-process a set $L_i$ of size $n_i$ for Algorithm 3-4, we need $\mathbf{O}(n_i \log n_i)$ time and $\mathbf{O}(n_i)$ space (in words).*

### 3.3 From Theory to Practice

In this section, we describe a more practical version of our methods. This algorithm is simpler, uses significantly less memory, straight-forward data structures, and, while it has worse theoretical guarantees, is faster in practice. The main difference is that – for each small group $L_i^z$, we only store the elements in $L_i^z$ and their images under $m \geq 1$ hash functions (i.e., we do not maintain inverted mappings, trading off a complex $\mathbf{O}(1)$-access for a simple scan over a short block of data). Also, we use only a single partition for each set $L_i$. Having multiple word representations of hash images (different hash functions) for each small group allows us to detect empty intersections of small groups with higher probability.

**Pre-processing Stage:** As before, each set $L_i$ is partitioned into groups $L_i^z$'s using a hash function $g_{t_i} : \Sigma \to \{0,1\}^{t_i}$. We will show that a good selection of $t_i$ is $\lceil \log(n_i/\sqrt{w}) \rceil$, which depends only on the size of $L_i$. Thus for each set $L_i$, pre-processing with a single partitioning suffices, saving significant memory. For each group, we compute word representations of images under $m$ (independent) universal hash functions $h_1, \ldots, h_m : \Sigma \to [w]$. Note that we only require a small value of $m$ in practice (e.g., $m = 2$).

**Online Processing Stage:** The algorithm for computing $\cap_i L_i$ we use here (Algorithm 5) is identical to Algorithm 4, with two exceptions: (1) When needed, $\cap_i L_i^{z_i}$ is directly computed by a linear merge of $L_i^{z_i}$'s (line 4), using $\mathbf{O}(\Sigma_i |L_i^{z_i}|)$ time. (2) We can skip the computation of $\cap_i L_i^{z_i}$ if, for *some* $h_j$, the bitwise-AND of the corresponding word representations $h_j(L_i^{z_i})$'s is zero (line 3).

---
1: **for** each $z_k \in \{0,1\}^{t_k}$ ($t_i = \lceil \log(n_i/\sqrt{w}) \rceil$) **do**
2:     Let $z_i$ be the $t_i$-prefix of $z_k$ for $i = 1, \ldots, k-1$
3:     **if** $\cap_{i=1}^k h_j(L_i^{z_i}) \neq \emptyset$ for all $j = 1, \ldots, m$ **then**
4:         Compute $\cap_{i=1}^k L_i^{z_i}$ by a linear merge of $L_1^z, \ldots, L_k^z$
5:         Let $\Delta \leftarrow \Delta \cup (\cap_{i=1}^k L_i^{z_i})$
6: $\Delta$ is the result of $\cap_{i=1}^k L_i$

---

**Algorithm 5:** "Simple" Intersection via Randomized Partitioning

**Analysis:** To see why Algorithm 5 is efficient, we observe that: if $L_1^{z_1} \cap L_2^{z_2} = \emptyset$, then with high probability, $h_j(L_1^{z_1}) \cap h_j(L_2^{z_2}) = \emptyset$ for some $j = 1, \ldots, m$. So most empty intersections can be skipped using the test in line 3. With the probability of a "*successful filtering*" (i.e. given $\cap_i L_i^{z_i} = \emptyset$, $\cap_i h_j(L_i^{z_i}) = \emptyset$ for some hash function $h_j$, $j = 1, \ldots, m$) bounded by the Lemmas A.1 and A.3, we can derive Theorem 3.9. Detailed analysis of this probability (both theoretical and experimental) and overall complexity is deferred to Appendix A.5.

THEOREM 3.9. *Using $t_i = \lceil \log(n_i/\sqrt{w}) \rceil$, Algorithm 5 computes $\cap_{i=1}^k L_i$ in expected $\mathbf{O}\left( \frac{\max(n, kn_k)}{\alpha(w)^m} + \frac{mn}{\sqrt{w}} + kr\sqrt{w} \right)$ time* $(r = |\cap_{i=1}^k L_i|, n = \sum_{i=1}^k n_i, \alpha(w) = \frac{1}{1-\beta(w)}$ *for $\beta(w)$ used in Lemma A.3).*

#### 3.3.1 Data Structure for Storing $L_i^z$

In this section, we describe the simple and space-efficient data structure that we use in Algorithm 5. As stated earlier, we only need to partition $L_i$ using one hash function $g_{t_i}$; hence we can represent each $L_i$ as an array of small groups $L_i^z$'s, ordered by $z$. For each small group, we store the information associated with it in the structure shown in Figure 3. The first word in this structure stores $z = g_{t_i}(L_i^z)$. The second word stores the structure's length $\mathsf{len}$. The following $m$ words represent the hash images $h_1(L_i^z), \ldots, h_m(L_i^z)$ of $L_i^z$. Finally, we store the elements of $L_i^z$ as an array in the remaining part. We need $n_i/\sqrt{w}$ such blocks for $L_i$ in total. The first word $z$ can be also computed on-the-fly, as these small groups are accessed sequentially in Algorithm 5. So, if we store len using one word, and one word for each element of $L_i^z$, then we need totally $m + 1 + |L_i^z|$ words for each group $L_i^z$, and thus $n_i(1 + (m+1)/\sqrt{w})$ words to store the pre-processed $L_i$. The overhead of the pre-processing is dominated by the cost of sorting $L_i$ (the remaining operations are trivial).

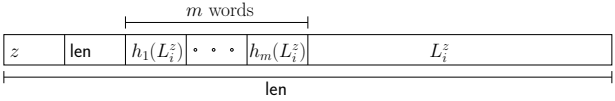

**Figure 3: The Structure for a Pre-processed Small Group $L_i^z$**

THEOREM 3.10. *To pre-process a set $L_i$ of size $n_i$ for Algorithm 5, we need $\mathbf{O}(n_i(m + \log n_i))$ time, and $\mathbf{O}(n_i(1 + m/\sqrt{w}))$ (words) space.*

We describe methods for compressing this structure in Appendix B.

## 3.4 Intersecting Small and Large Sets

An important special case for set intersection are asymmetric intersections where the sizes $n_1$ and $n_2$ of the sets that are intersected vary significantly (w.l.o.g., assume $n_1 \ll n_2$). In this subsection, using the same multi-resolution data structure as in Section 3.2.1, we present an algorithm HashBin that computes $L_1 \cap L_2$ in $\mathbf{O}(n_1 \log(n_2/n_1))$ time. This bound is also achieved by other previous works, e.g., SmallAdaptive [5], but our algorithm is even simpler in online processing. It is also known that algorithms based on hash-tables only require $\mathbf{O}(n_1)$ time for this scenario; however, unlike HashBin, they are ill-suited for less asymmetric cases.

**Algorithm HashBin:** When intersecting two sets $L_1$ and $L_2$ with sizes $n_1 \ll n_2$, we focus on the partitioning induced by $g_t : \Sigma \to \{0, 1\}^t$, where $t = \lceil \log n_1 \rceil$ for both of them, and $g$ is a random permutation of $\Sigma$. To compute $L_1 \cap L_2$, we compute $L_1^z \cap L_2^z$ for all $z \in \{0, 1\}^t$ and take the union. To compute $L_1^z \cap L_2^z$, we iterate over each $x \in L_1^z$, and perform a binary search to check whether $x \in L_2^z$ using $\mathbf{O}(\log |L_2^z|)$ time. This scheme can be extended to multiple sets by searching for $x$ in $L_i^z$ if found in $L_1^z, \ldots, L_{i-1}^z$.

THEOREM 3.11. *The algorithm HashBin computes $L_1 \cap L_2$ in expected $\mathbf{O}\left(n_1 \log \frac{n_2}{n_1}\right)$ time. The pre-processing of a list $L_i$ requires $\mathbf{O}(n_i \log n_i)$ time and $\mathbf{O}(n_i)$ space.*

The proof of Theorem 3.11 and how HashBin uses the multi-resolution data structure is deferred to the Section A.6 in the appendix. The advantage of HashBin is that, since it is based on the same structure as the algorithm introduced in Section 3.2, we can make the choice between algorithms online, based on $n_1/n_2$.

## 4. EXPERIMENTAL EVALUATION

We evaluate the performance and space requirements of four of the techniques described in this paper: (a) the fixed-width partition algorithm described in Section 3.1 (which we will refer to as IntGroup); (b) the randomized partition algorithm in Section 3.2 (RanGroup) (c) the simple algorithm based on randomized partitions described in Section 3.3 (RanGroupScan); and (d) the one for intersecting sets of skewed sizes in Section 3.4 (HashBin).

**Setup:** All algorithms are implemented using C and evaluated on a 4GB 64-bit 2.4GHz PC. We employ a random permutation of the document IDs for the hash function $g$ and 2-universal hash functions for $h$ (or $h_j$'s). For RanGroup, we use $m = 4$ (the number of hash functions $h_j$), unless noted otherwise.

We compare our techniques to the following competitors: (i) set intersection based on a simple parallel scan of inverted indexes: Merge; (ii) set intersection based on skip lists: SkipList [18]; (iii) set intersection based on hash tables: Hash (i.e., we iterate over the smallest set $L_1$, looking up every element $x \in L_1$ in hash-table representations of $L_2, \ldots L_k$); (iv) the algorithm of [6]: BPP; (v) the algorithm proposed for fast intersection in integer inverted indices in main memory [19, 21]: Lookup (using $B = 32$ as the bucket-size, which is the best value in our and the authors' experience); and (vi) various *adaptive* intersection algorithms based on binary search/galloping search: SvS, Adaptive [12, 13, 3], BaezaYates [1, 2], and SmallAdaptive [5]. Note that BaezaYates is generalized to handle more than two sets as in [5].

**Implementation:** For each competitor, we tried our best to optimize its performance. For example, for Merge we tried to minimize the number of branches in the inner loop; we also store postings in consecutive memory addresses to speed up parallel scans and reduce page walks after TLB misses. Our implementation of skip lists follows [18], with simplifications since we are focusing on static data and do not need fast insertion/deletion. We also simplified the bit-manipulation in BPP [6] so that it works faster in practice for small $w$. For the algorithms using inverted indexes, we initially do not consider compression on the posting lists, as we do not want the decompression step to impact the performance reports. In Section 4.1 we will study variants of the algorithms incorporating compression. With regards to skip-operations in the index note that since we use uncompressed posting lists, algorithms such as Adaptive can perform arbitrary skips into the index directly.

**Datasets:** To evaluate these algorithms we use both synthetic and real data. For the experiments with synthetic datasets, sets are generated randomly (and uniformly) from a universe $\Sigma$. The real dataset is a collection of more than 8M Wikipedia pages. In each experiment for the synthetic datasets, 20 combinations of sets are randomly generated, and the average time is reported.

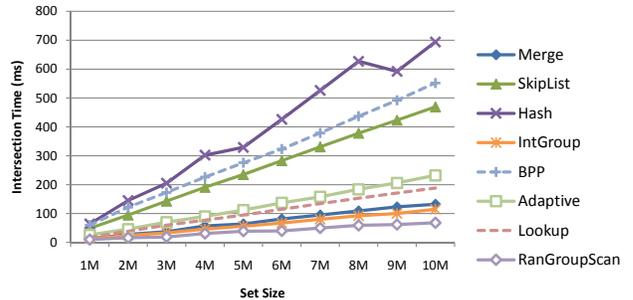

**Figure 4: Varying the Set Size**

**Varying the Set Size:** First, we measure the performance when intersecting only 2 sets; we use synthetic data, the lists are of equal size and the size of the intersection is fixed at 1% of the list size; the results are shown in Figure 4. We can see that the performance of the different techniques relative to each other does not change with varying list size. Hash performs worst, as the (relatively) expensive lookup operation needs to be performed many times. SkipList performs poorly for the same reason. The BPP algorithm is also slow, but this is because of a number of complex operations that need to be performed, which are hidden as a constant in the $\mathbf{O}()$-notation. The same trend held for the remaining experiments as well; hence, for readability, we did not include BPP in the subsequent graphs. For the same reason we only show the best-performing among the *adaptive* algorithms in the evaluation; if one adaptive algorithm dominates another on all parameter settings in an experiment, we don't plot the worse one.

Among the remaining algorithms, RanGroupScan (40%-50% faster than Merge) and IntGroup perform the best (RanGroup performs similarly to IntGroup and is not plotted). Interestingly,

260

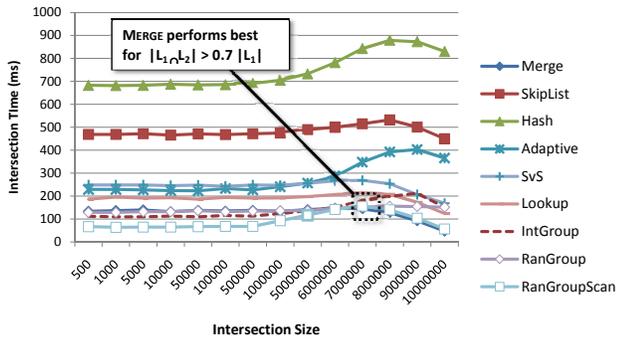

Figure 5: **Varying the Intersection Size**

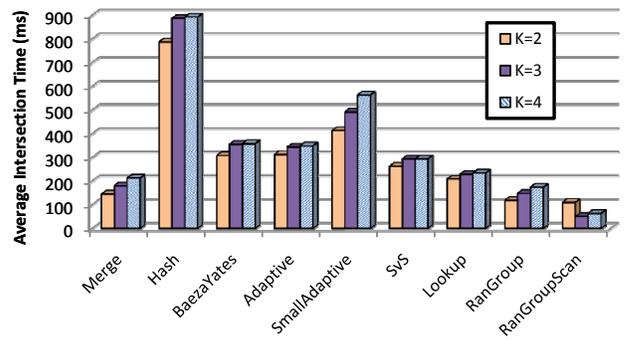

Figure 6: **Varying the Number of Keywords**

the simple Merge algorithm is next, outperforming the more sophisticated algorithms, followed by Lookup and the best-performing *adaptive* algorithm.

**Varying the Intersection Size:** The size of the intersection $r$ is an important factor concerning the performance of the algorithms: larger intersections mean fewer opportunities to eliminate small groups early for our algorithms or to skip parts of the set for the adaptive and skiplist-based approaches. Here, we use synthetic data, intersecting two sets with 10M elements and vary $r = |L_1 \cap L_2|$ between 500 and 10M. The results are reported in Figure 5. For $r < 7M$ (70% of the set size) RanGroupScan and IntGroup perform best. Otherwise, Merge becomes the fastest and RanGroupScan the 2nd-fastest alternative; here, the performance of RanGroupScan is very similar to Merge, all the way to $r = 10M$. Among the remaining algorithms, RanGroup slightly outperforms Merge for $r < 5M$, Lookup is the next-best algorithm and SvS and Adaptive perform best among the *adaptive* algorithms.

**Varying the Sets Size Ratios:** As we illustrated in the introduction, the skew in set sizes is also an important factor in performance. When sets are very different in size, algorithms that iterate through the smaller set and are able to locate the corresponding values in the larger set quickly, such as HashBin and Hash, perform well. In this experiment we use synthetic data and vary the ratio of set sizes, setting $|L_2| = 10M$ and varying $|L_1|$ between 16K and 10M. The size of the intersection is set to be 1% of $|L_1|$ and we define the ratio between the list sizes as $sr = |L_2|/|L_1|$. Here, the differences between the algorithms become small with growing $sr$ (for this reason, we also don't report them in a graph, as too many lines overlap). For $sr < 32$, RanGroupScan performs best; for larger $sr$, Lookup and Hash perform best, until a ratio of $sr \geq 100$ – for this and larger ratios, Hash outperforms the remaining algorithms, followed by Lookup and HashBin. Generally, both HashBin and RanGroupScan perform close to the best-performing algorithm. The adaptive algorithms require more time than RanGroupScan for $sr \leq 200$ and more time than HashBin for all values of $sr$; Skiplist and BPP perform worst across all values of $sr$.

**Varying the Number of Keywords:** In this experiment, we varied the number of sets $k = 2, 3, 4$, fixing $|L_i| = 10M$ for $i = 1, \ldots, k$, with the IDs in the sets being randomly generated using a uniform distribution over $[0, 2 \times 10^8]$; the results are reported in Figure 6. In this experiment, we use $m = 2$ hash images for RanGroupScan. For multiple sets, RanGroupScan is the fastest, with the difference becoming more pronounced for 3 and 4 keywords, since, with additional sets, intersecting the hash-images (word-representations) yields more empty results, allowing us to skip the corresponding groups. RanGroup is the next-best performing algorithm; we don't include results for IntGroup here, as it is designed for intersections of two sets (see Section 3.1). Interestingly, the simple Merge algorithm again performs very well when compared to the more sophisticated techniques; the Lookup algorithm is next, followed by the various *adaptive* techniques.

**Size of the Data Structure:** The improvements in speed come at the cost of an increase in space: our data structures (without compression) require more space than an uncompressed posting list – the increase is 37% (RanGroupScan for $m = 2$), 63% (RanGroupScan for $m = 4$), 75% (IntGroup) or 87% (RanGroup).

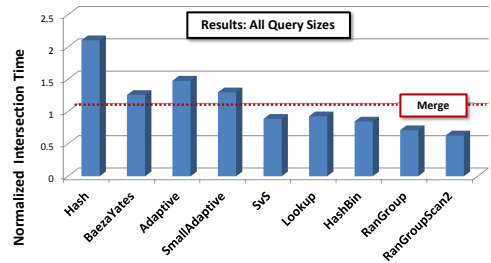

Figure 7: **Normalized Execution Time on a Real Workload**

**Experiment on Real Data:** In this experiment, we used a "workload" of the $10^4$ most frequent (measured over a week in 2009) queries against the *Bing.com* search engine. As the text corpus, we used a set of 8 Million Wikipedia documents.
*Query characteristics:* 68% of the queries contain 2 keywords, 23% 3 keywords and 6% 4 keywords. As we have illustrated before, a key factor for performance is the ratio of set sizes – among the 2-word queries, the average ratio $|L_1|/|L_2|$ is 0.21, for 3-word queries the average ratio $|L_1|/|L_2|$ is 0.31 and the average ratio $|L_1|/|L_3|$ is 0.09, and for 4-word queries, the $|L_1|/|L_2|$ ratio is 0.36 and the $|L_1|/|L_4|$ ratio is 0.06 – note that $|L_1| \leq |L_2| \leq |L_3| \leq |L_4|$. The average ratio of intersection size to $|L_1|$ is 0.19.

To illustrate the relative performance of the algorithms over all queries we plotted their average running times in Figure 7: here, the running time of Merge is normalized to 1. Both RanGroup and RanGroupScan significantly outperform Merge, with the latter performing the best overall; interestingly, when used for all queries (as opposed to only for the large skew case it was designed for) HashBin still performed better than Merge. The remaining algorithms performed in similar order to the earlier experiments, with the one exception being SvS which outperformed both Merge and Lookup for this more realistic data. Overall, the RanGroupScan was the best-performing algorithm for 61.6% of the queries, followed by RanGroup (16%) and Hashbin (7.7%) – among the remaining algorithms not proposed in this paper, Lookup performed best in 6.4% of the queries and SvS for 3.6% of the queries. All of the other techniques were best for 2.1% of the queries or fewer. We present additional experiments for this data set in the Appendix C.2.

## 4.1 Experiments on Compressed Structures

To illustrate the impact of compression on performance, we repeated the first experiment above, intersecting two sets of identical size, with the size of the intersection fixed to 1% of the set size. Varying the set size, we report the execution times and storage requirements for the three algorithms that performed best overall in the earlier experiments – Merge, Lookup and RanGroupScan (since we are interested in small structures here, we only use $m = 1$ hash images in RanGroupScan) – when being compressed with different techniques: we used the standard techniques based on $\gamma$- and $\delta$-coding (see [23], p.116) to compress the parts of the posting data stored and accessed sequentially for the three algorithms, and the compression technique described in Appendix B for RanGroupScan (which we refer to as RanGroupScan_Lowbits). The results are shown in Figure 8; here, we omitted the results for $\gamma$-encoding as they were essentially indistinguishable from ones for $\delta$-coding. RanGroupScan outperforms – in terms of speed – the other two algorithms using the same compression scheme; the other two algorithms perform similarly to each other, as the decompression now dominates their run-time. Using our encoding scheme of Appendix B improves the performance significantly.

Looking at the graph, we can see that the storage requirement for RanGroupScan (using our own encoding) is between 1.3-1.9x of the size of the compressed inverted index and between 1.2-1.6x of the compressed Lookup structure. At the same time, the performance improvements are between 7.6-15x (vs. Merge) or 7.4-13x (vs. Lookup). Furthermore, by increasing the number of hash images to $m = 2$, we obtain an algorithm that significantly outperforms the *uncompressed* Merge, while requiring less memory.

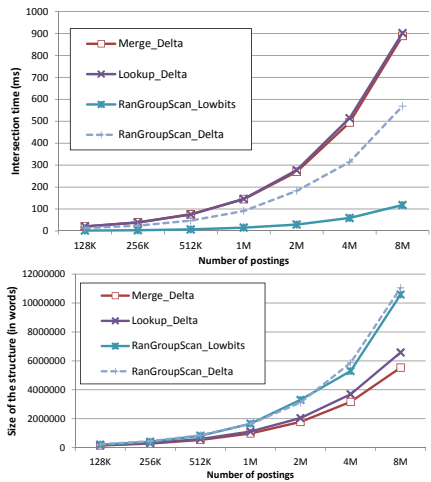

**Figure 8:** Running Time and Space Requirement

**Experiment on Real Data:** We repeated this experiment using the real-life data/workload described earlier and the compressed variants of RanGroupScan, Lookup and Merge. Again, RanGroupScan_Lowbits performed best, improving run-times by a factor of 8.4x (vs. Merge + $\delta$-coding), 9.1x (Merge + $\gamma$-coding), 5.7x (Lookup + $\delta$-coding), 6.2x (Lookup + $\gamma$-coding), respectively. However, our approach required the most space (66% of the uncompressed data), whereas Merge (26% / 28% for $\gamma$- / $\delta$-coding) and Lookup (35% / 37%) required significantly less.

Finally, to illustrate the robustness of our techniques, we also measured the worst-case latency for any single query: here, the worst-case latency using Merge + $\delta$-coding was 5.2x the worst-case latency of RanGroupScan_Lowbits. We saw similar results for Merge + $\gamma$-coding (5.6x higher), Lookup + $\delta$-coding (4.4x higher), and Lookup + $\gamma$-coding (4.9x higher).

## 5. CONCLUSION

In this paper we introduced algorithms for set intersection processing for memory-resident data. Our approach provides both novel theoretical worst-case guarantees as well as very fast performance in practice, at the cost of increased storage space. Our techniques outperform a wide range of existing techniques and are robust in that – for inputs for which they are not the best-performing approach – they perform close to the best one. Our techniques have applications in information retrieval and query processing scenarios where performance is of greater concern than space.

# APPENDIX

## Acknowledgments

We thank the anonymous reviewers for their numerous insights and suggestions that immensely improved the paper.

## A. PROOFS OF THEOREMS

### A.1 Analysis of Algorithm 1 (Proof of Theorem 3.3)

There are a total of $\mathbf{O}((n_1+n_2)/\sqrt{w})$ pairs of $L_1^p$ and $L_2^q$ to be checked in Algorithm 1. For each pair, since $H = h(L_1^p) \cap h(L_2^q)$ can be computed in $\mathbf{O}(1)$ time and elements in $H$ can be enumerated in linear time, the cost of computing $L_1^p \cap L_2^q$ is dominated by computing $h^{-1}(y, L_1^p) \cap h^{-1}(y, L_2^p)$ for every $y \in H$, the cost of which is in turn determined by the number of pairs of elements which are mapped to the same location by $h$; we denote this set as $I = \{(x_1, x_2) \mid x_1 \in L_1^p, x_2 \in L_2^q, \text{ and } h(x_1) = h(x_2)\}$.

Let $I_= = \{(x_1, x_2) \mid x_1 = x_2\} \cap I$ denote the pairs of identical elements (i.e., elements in the intersection) in $I$ and $I_{\neq} = \{(x_1, x_2) \mid x_1 \neq x_2\} \cap I$ the remaining pairs of elements that are hashed to the same value by $h$ but are not identical. Obviously, $|I_=| = |L_1^p \cap L_2^q|$. If we can show $\mathbf{E}[|I_{\neq}|] = \mathbf{O}(1)$, the proof is completed: this is because, for a total of $\mathbf{O}((n_1+n_2)/\sqrt{w})$ pairs of $L_1^p$ and $L_2^q$ to be checked, the total running time is

$$\sum_{p,q} \mathbf{O}(\mathbf{E}[|I|]+1) = (\sum_{p,q}|I_=| + \sum_{p,q}(\mathbf{E}[|I_{\neq}|]+1)) \cdot \mathbf{O}(1)$$
$$= \mathbf{O}(r) + (n_1+n_2)/\sqrt{w} \cdot \mathbf{O}(1). \quad (3)$$

Indeed, we can show for each pair of $L_1^p$ and $L_2^q$ that:

$$\mathbf{E}[|I_{\neq}|] = \sum_{\substack{x_1 \in L_1^p, x_2 \in L_2^q, \\ x_1 \neq x_2}} \mathbf{Pr}[h(x_1) = h(x_2)] \leq \sqrt{w} \cdot \sqrt{w} \cdot \frac{1}{w} = \mathbf{O}(1), \quad (4)$$

for a universal hash function $h$, which completes the proof. $\square$

#### A.1.1 Group Size and Optimizing Running Time

In Algorithm 1, the group size is selected as the "magical number" $\sqrt{w}$ (i.e., $|L_1^p| = |L_2^q| = \sqrt{w}$). To explain this choice, we now explore the effect of group size on the running time of Algorithm 1. Suppose in general $L_i$ is partitioned into groups of size $s_i$. Extending Equation (4) a bit, we have $\mathbf{E}[|I_{\neq}|] = \mathbf{O}(1)$ as long as $s_1 \cdot s_2 \leq w$. Then following the same argument as in (3), a total of $\mathbf{O}(n_1/s_1 + n_2/s_2)$ pairs are to be checked, and the expected running time of Algorithm 1 is $\mathbf{O}(T(s_1, s_2))$, where $T(s_1, s_2) = n_1/s_1 + n_2/s_2 + r$. Minimizing $T(s_1, s_2)$ under the constraint $s_1 \cdot s_2 \leq w$ yields optimal group sizes of $s_1^* = \sqrt{wn_1/n_2}$ and $s_2^* = \sqrt{wn_2/n_1}$, and the optimal running time is $\mathbf{O}(T(s_1^*, s_2^*)) = \mathbf{O}(\sqrt{n_1 n_2/w} + r)$. If we now use the group sizes $s_1' = s_2' = \sqrt{w}$, as in the proof of Theorem 3.3, we obtain a running time of $\mathbf{O}(T(s_1', s_2')) = \mathbf{O}((n_1+n_2)/\sqrt{w} + r)$.

$\mathbf{O}(\sqrt{n_1 n_2/w} + r)$ is better than $\mathbf{O}((n_1+n_2)/\sqrt{w} + r)$ when set sizes are skewed (e.g., $n_1 \ll n_2$ or $n_1 = \sqrt{n_2}$). To achieve the better bound we leverage that the group size $s_1^* = \sqrt{wn_1/n_2}$ of the set $L_1$ depends on the size $n_2$ of the set $L_2$ to be intersected with it, and use a multi-resolution structure which keeps different partitions of a set, as discussed at the end of Section 3.1.

### A.2 Analysis of Algorithm 3 (Proof of Theorem 3.5)

Similar to the proof of Theorem 3.3, the cost of computing $L_1^{z_1} \cap L_2^{z_2}$ using IntersectSmall for each pair of small groups $L_1^{z_1}$ and $L_2^{z_2}$ is determined by the size of $I = \{(x_1, x_2) \mid x_1 \in L_1^{z_1}, x_2 \in L_2^{z_2}, \text{ and } h(x_1) = h(x_2)\}$. As in A.1, let $I_= = \{(x_1, x_2) \mid x_1 = x_2\} \cap I$ and $I_{\neq} = \{(x_1, x_2) \mid x_1 \neq x_2\} \cap I$. Obviously, $|I_=| = |L_1^{z_1} \cap L_2^{z_2}|$, and $I_{\neq}$ is the set of element-pairs that result in a hash-collision. If we can show $\mathbf{E}[|I_{\neq}|] \leq \mathbf{O}(1)$, the proof is complete: because, since $t_1 = t_2 = \lceil \log \sqrt{n_1 n_2/w} \rceil$, there are $\mathbf{O}(\sqrt{n_1 n_2/w})$ pairs of $z_1$ and $z_2$ to be considered (we have $z_1 = z_2 = z$ in every iteration), and thus the total running time is

$$\sum_{z \in \{0,1\}^{t_2}} \mathbf{O}(\mathbf{E}[|I|]+1) = (\sum_{z \in \{0,1\}^{t_2}} |I_=| + \sum_{z \in \{0,1\}^{t_2}} (\mathbf{E}[|I_{\neq}|]+1)) \cdot \mathbf{O}(1)$$
$$\leq \mathbf{O}(r) + \sqrt{n_1 n_2}/\sqrt{w} \cdot \mathbf{O}(1).$$

Now we prove that, for each pair $(z_1, z_2)$, $\mathbf{E}[|I_{\neq}|] = \mathbf{O}(1)$. Letting $S_1^{z_1} = |L_1^{z_1} - L_2^{z_2}|$ and $S_2^{z_2} = |L_2^{z_2} - L_1^{z_1}|$, if $L_1^{z_1}$ and $L_2^{z_2}$ are fixed, we have (similar to (4) in the proof of Theorem 3.3):

$$\mathbf{E}_h[|I_{\neq}| \mid S_1^{z_1}, S_2^{z_2}] = S_1^{z_1} \cdot S_2^{z_2}/w.$$

$S_1^{z_1}$ and $S_2^{z_2}$ are random variables determined by the hash function $g$. From their definition and the property of 2-universal (2-independent) hashing, we can prove $\mathbf{E}_g[S_1^{z_1} \cdot S_2^{z_2}] \leq \mathbf{E}_g[S_1^{z_1}] \cdot \mathbf{E}_g[S_2^{z_2}]$ (using a random permutation $g$ yields the same result). Also, $\mathbf{E}_g[S_1^{z_1}] \leq \mathbf{E}_g[|L_1^{z_1}|] = \mathbf{O}(n_1/2^{t_1}) = \mathbf{O}(\sqrt{n_1 w/n_2})$, and similarly $\mathbf{E}_g[S_2^{z_2}] \leq \mathbf{O}(\sqrt{n_2 w/n_1})$. Therefore, $\mathbf{E}_g[S_1^z \cdot S_2^z] = \mathbf{E}_g[S_1^z] \cdot \mathbf{E}_g[S_2^z] \leq \mathbf{O}(w)$, and thus, $\mathbf{E}[|I_{\neq}|] =$

$$= \mathbf{E}_g[\mathbf{E}_h[|I_{\neq}| \mid S_1^z, S_2^z]] = \mathbf{E}_g\left[\frac{S_1^z \cdot S_2^z}{w}\right] \leq \frac{\mathbf{O}(w)}{w} = \mathbf{O}(1), \quad (5)$$

which completes the argument. $\square$

### A.3 Analysis of Algorithm 4 (Proof of Theorem 3.6 and Theorem 3.7)

Theorem 3.6 is special case of Theorem 3.7 for two-set intersection. So we only present the proof of Theorem 3.7 below.

Consider any element $x \in L_i$ for each set $L_i$ involved in the intersection computation, i.e., extended IntersectSmall in line 3 of Algorithm 4, where we compute:

$$H = \bigcap_{i=1}^k h(L_i^{z_i}), \text{ and } \bigcap_{i=1}^k L_i^{z_i} = \bigcup_{y \in H}\left(\cap_{i=1}^k h^{-1}(y, L_i^{z_i})\right).$$

Denote the set of all such elements (with $h(x) = y \in H$) by $\Gamma$. The number of such elements $|\Gamma|$ dominates the cost of Algorithm 4. We first differentiate two cases of elements in $\Gamma$:

(i) $x \in \bigcap_{i=1}^k L_i$: These $r$ elements are scanned $k$ times, and thus contribute a factor of $\mathbf{O}(kr)$ in the time complexity overall.

(ii) $x \notin \bigcap_{i=1}^k L_i$: We group all these elements into $k-1$ sets, $D_2, \ldots, D_k$ (an element $x$ may belong to multiple $D_i$'s):

$$D_i = \{x \in \Gamma \mid x \in L_i \cap L_{i+1} \cap \ldots \cap L_k \wedge x \notin L_j \text{ for some } j < i\}.$$

Now focus on $D_i \cap L_i^{z_i}$ for each $z_i \in \{0,1\}^{t_i}$. For any $x \in L_i$ but $x \notin L_j$ for some $j < i$, letting $z_j$ be the $t_j$-prefix of $z_i$, we have $x \in D_i \cap L_i^{z_i}$ implies that $h(x) \in H$ and thus there exists $x'(\neq x) \in L_j^{z_j}$ such that $h(x) = h(x')$; so for such an $x$,

$$\mathbf{Pr}[x \in D_i \cap L_i^{z_i} \mid x \in L_i^{z_i}] = \mathbf{Pr}[h(x) \in H \mid x \in L_i^{z_i}]$$
$$\leq \sum_{x' \in L_j^{z_j}} \mathbf{Pr}[h(x) = h(x')] \leq |L_j^{z_j}|/w.$$

Generalizing Equation (5) in the proof of Theorem 3.5, we have

$$\mathbf{E}\left[|D_i \cap L_i^{z_i}|\right] = \mathbf{E}_g\left[\mathbf{E}_h\left[|D_i \cap L_i^{z_i}| \mid |L_j^{z_j}| \text{ for all } z_j\text{'s}\right]\right]$$

$$\leq \sum_{x \in L_i} \mathbf{E}_g\left[|L_j^{z_j}|/w\right] \cdot \mathbf{Pr}\left[x \in L_i^{z_i}\right] \leq n_i \cdot \frac{\sqrt{w}}{w} \cdot \frac{\sqrt{w}}{n_i} = \mathbf{O}(1)$$

(as $\mathbf{E}_g\left[|L_j^{z_j}|\right] = \sqrt{w}$ for any $j$, and $\mathbf{Pr}\left[x \in L_i^{z_i}\right] = \sqrt{w}/n_i$). So $\mathbf{E}\left[|D_i|\right] \leq \mathbf{O}(n_i/\sqrt{w})$, as $L_i$ is partitioned into $2^{t_i} = n_i/\sqrt{w}$ groups $L_i^{z_i}$'s over all iterations of Algorithm 4. Then we have

$$\mathbf{E}\left[\sum_{i=2}^{k} |D_i|\right] \leq \sum_{i=2}^{k} \mathbf{O}(n_i/\sqrt{w}) = \mathbf{O}(n/\sqrt{w}). \quad (6)$$

**Running Time:** As the $|D_i|$'s are bounded as above, a naive implementation of Algorithm 4 requires $\mathbf{O}(kn_k/\sqrt{w} + kr)$ time in expectation. The iteration of lines 1-4 in Algorithm 4 repeats $n_k/\sqrt{w}$ times (suppose $n_1 \leq n_2 \leq \ldots \leq n_k$). In each iteration, we compute $H$ in $O(k)$ time, and each element in $D_k$ needs $\mathbf{O}(k)$ comparisons to be eliminated.

$n/\sqrt{w}$ is potentially smaller than $kn_k/\sqrt{w}$ especially for sets with skewed sizes. With careful memorization of the partial results $\bigcap_{i=1}^{j} h(L_i^{z_i})$ and $\bigcap_{i=j}^{k} h^{-1}(y, L_i^{z_i})$ in Algorithm 4, from (i) and (ii), we now prove the promised running time $\mathbf{O}(n/\sqrt{w} + kr)$:

The major cost of Algorithm 4 comes from the computation of (a) $H = \bigcap_{i=1}^{k} h(L_i^{z_i})$ and (b) $\bigcap_{i=1}^{k} h^{-1}(y, L_i^{z_i})$ for each $y \in H$. Assume $n_1 \leq n_2 \leq \ldots \leq n_k$.

For (a), as $z_i$ is the $t_i$-prefix of $z_j$ if $i \leq j$, we can memorize $\bigcap_{i=1}^{j} h(L_i^{z_i})$ for each $z_j$. Then, for example, reuse $h(L_1^1) \cap h(L_2^{10})$ when computing $h(L_1^1) \cap h(L_2^{10}) \cap h(L_3^{100})$ and $h(L_1^1) \cap h(L_2^{10}) \cap h(L_3^{101})$. In this way, the computation of $H$ for different combinations of $z_1, \ldots, z_k$ requires $\sum_i \mathbf{O}(n_i/\sqrt{w}) = \mathbf{O}(n/\sqrt{w})$ time.

For (b), for each combination of $z_1, \ldots, z_k$, we compute the result the inverse order (from $i = k$ to $i = 1$): the partial results $\bigcap_{i=j}^{k} h^{-1}(y, L_i^{z_i})$ (for all $y \in H$, all $z_k$'s, and some $j$) have their total size bounded by $|D_j| + r$. Using the hash-table-based approach to compute the intersection, the total running time is bounded by the total size of the partial results. So from (6), the total running time is $\mathbf{O}(n/\sqrt{w} + kr)$ in expectation. $\square$

## A.4 Analysis of the Multi-resolution Structure (Proof of Theorem 3.8)

The time bound is trivial, because we only require sorting and scanning of each set. The total space for storing $h(L_i^z)$'s, left$(L_i^z)$'s, and right$(L_i^z)$'s is $\mathbf{O}(n_i)$, as there are $\mathbf{O}(n_i + n_i/2 + n_i/4 + \ldots) = \mathbf{O}(n_i)$ groups of different sizes. For next$(x)$'s we also only need $\mathbf{O}(n_i)$ space, as there are $n_i$ elements in the set. We now analyze the space needed for first$(y, L_i^z)$'s to complete the proof.

To store first$(y, L_i^z)$ for each $y \in [w]$ and each $z$, storing the difference between first$(y, L_i^z)$ and left$(L_i^z)$ suffices; so we need $\mathbf{O}(\log |L_i^z|)$ *bits*. To store first$(y, L_i^z)$'s for all $y \in [w]$ in a group $L_i^z$, we need $\mathbf{O}(w \cdot \log |L_i^z|/w) = \mathbf{O}(\log |L_i^z|)$ *words*. Consider the partitioning induced by $g_t : \Sigma \to \{0, 1\}^t$ for some $t$, letting $\bar{t} = \lceil \log n_i \rceil - t$, there are $\mathbf{O}(n_i/2^{\bar{t}})$ groups $L_i^z$'s generated by $g_t$, so the space we need for all these groups is: $(\log(\cdot)$ is concave)

$$\mathbf{O}(\sum_{z \in \{0,1\}^t} \log |L_i^z|) \leq \mathbf{O}(2^t \cdot \log(n_i/2^t)) = \mathbf{O}((n_i/2^{\bar{t}}) \cdot \bar{t}).$$

Therefore, for all resolutions $t = 1, 2, \ldots, \lceil \log n_i \rceil$, the total space needed for first$(y, L_i^z)$'s is $\mathbf{O}(\sum_{\bar{t}} \bar{t} \cdot n_i/2^{\bar{t}}) \leq \mathbf{O}(n_i)$. $\square$

## A.5 Analysis of Algorithm 5 (Proof of Theorem 3.9)

### A.5.1 Probability of Successful Filtering

Recall in Algorithm 5, sets are partitioned into small groups by hash function $g$, and $m$ universal hash functions $h_1, \ldots, h_m$ are used to test whether the intersection of small groups is empty.

It is efficient because of the following observation: if $L_1^{z_1} \cap L_2^{z_2} = \emptyset$, then $h_j(L_1^{z_1}) \cap h_j(L_2^{z_2}) = \emptyset$ for some $j = 1, \ldots, m$ (so-called "*successful filtering*") with high probability. But once a "*false positive*" happens (i.e., $\cap_i L_i^{z_i} = \emptyset$ but $\cap_i h_j(L_i^{z_i}) \neq \emptyset$ for any hash function $h_1, \ldots, h_m$), we have to scan the two or $k$ small groups for the intersection. So to analyze Algorithm 5, the key point we need to establish is that *successful filtering* happens with a constant probability for two or $k$ small groups. We first verify the above intuition, by assuming that $|L_i^{z_i}|$ is $\sqrt{w}$:

LEMMA A.1. *For two small groups $L_1^{z_1}$ and $L_2^{z_2}$ with $|L_1^{z_1}| = |L_2^{z_2}| = \sqrt{w}$, given a universal hash function $h : \Sigma \to [w]$, if $L_1^{z_1} \cap L_2^{z_2} = \emptyset$, then $h(L_1^{z_1}) \cap h(L_2^{z_2}) = \emptyset$ with probability at least $\left(1 - \frac{1}{\sqrt{w}}\right)^{\sqrt{w}}$ ($\approx 0.3436$ for $w = 64$).*

PROOF. Since $L_1^{z_1} \cap L_2^{z_2} = \varnothing$, for each $x_2 \in L_2^{z_2}$, we have $h(x_2) \notin h(L_1^{z_1})$ holds with probability $1 - |h(L_1^{z_1})|/w$. So,

$$\mathbf{Pr}\left[h(L_1^z) \cap h(L_2^z) = \emptyset\right] \geq \left(1 - \frac{|h(L_1^{z_1})|}{w}\right)^{|L_2^{z_2}|} \geq \left(1 - \frac{|L_1^{z_1}|}{w}\right)^{|L_2^{z_2}|},$$

as $|h(L_1^{z_1})| \leq |L_1^{z_1}|$. So, when $|L_1^{z_1}| = |L_2^{z_2}| = \sqrt{w}$, we have $\mathbf{Pr}\left[h(L_1^z) \cap h(L_2^z) = \emptyset\right] \geq \left(1 - \frac{1}{\sqrt{w}}\right)^{\sqrt{w}}$. $\square$

In general, although the sizes of the small groups $|L_i^z|$ are random variables, they are unlikely to deviate from $\sqrt{w}$ by much. This is important since groups of larger sizes result in poorer filtering performance of the word representations $h_j(L_i^{z_i})$'s (incurring more false positives). Using *Chernoff bounds* we can show that:

PROPOSITION A.2. *For any group $L_i^{z_i}$ defined in Algorithm 5 (i.e., partition $L_i$ by $g_{t_i} : \Sigma \to \{0, 1\}^{t_i}$ with $t_i = \lceil \log(n_i/\sqrt{w}) \rceil$), we have: (i) $\frac{\sqrt{w}}{2} \leq \mathbf{E}\left[|L_i^{z_i}|\right] \leq \sqrt{w}$;*
*(ii) $\mathbf{Pr}\left[|L_i^{z_i}| \leq (1+\epsilon)\sqrt{w}\right] \geq 1 - \exp\left(-\frac{\sqrt{w}\epsilon^2}{3}\right)$, for $0 < \epsilon < 1$;*
*(iii) $\mathbf{Pr}\left[|L_i^{z_i}| \leq \delta(w)\sqrt{w}\right] \geq 1 - \frac{1}{4\sqrt{w}}$, where the constant $\delta(w) = 1 + \left(\frac{6\ln(4\sqrt{w})}{\sqrt{w}}\right)^{1/2}$ ($\approx 2.6122$ for $w = 64$).*

PROOF. In this proof we use a random permutation as $g : \Sigma \to \Sigma$, and define $g_t(x)$ to be the $t$ most significant bits of $g(x)$. However, note that we can use a hash function here as well, if we use the pre-image to break any ties resulting from hash collisions (thereby resulting in a total ordering).

For the group $L_i^{z_i}$, define $Y_x = 1$ if $x \in L_i^{z_i}$ (i.e., $g_{t_i}(x) = z_i$), and $Y_x = 0$ otherwise. So $|L_i^{z_i}| = \sum_x Y_x$. Then (i) is from the fact that $\mathbf{Pr}\left[Y_x = 1\right] = 1/2^{t_i}$ and the linearity of expectation.

For a random permutation $g$, we can prove that the $\{Y_x \mid x \in L_i\}$ are *negatively associated* [14], so the Chernoff bounds can be still applied. As in (i), we have $\mu_L = \sqrt{w}/2 \leq \mu = \mathbf{E}\left[|L_i^{z_i}|\right] \leq \sqrt{w} = \mu_H$. To prove (iii), we use the Chernoff bound:
$\mathbf{Pr}\left[|L_i^{z_i}| > (1+\epsilon)\mu\right] < \exp(-\mu\epsilon^2/3) \leq \exp(-\mu_L\epsilon^2/3)$ [14].
To prove (ii), we can use a tighter bound: for $0 < \epsilon < 1$,
$\mathbf{Pr}\left[|L_i^{z_i}| > (1+\epsilon)\mu_H\right] < \exp(-\mu_H\epsilon^2/3)$ [14]. $\square$

Note that the same bounds hold when a hash function is used as $g$. Lemma A.3 extends Lemma A.1 for $k$ groups, whose sizes are random variables determined by the hash function $g$.

LEMMA A.3. *For $k$ groups $L_i^{z_i}$'s (for $i = 1, \ldots, k$, partition $L_i$ by $g_{t_i} : \Sigma \to \{0,1\}^{t_i}$ where $t_i = \lceil \log(n_i/\sqrt{w}) \rceil$), if $\cap_i L_i^{z_i} = \emptyset$, then $\cap_i h(L_i^{z_i}) = \emptyset$ with at least constant probability*
$$\beta(w) = \left(1 - \frac{1 + \delta(w)\sqrt{w}}{4\sqrt{w}}\right) \cdot \left(1 - \frac{\delta(w)}{\sqrt{w}}\right)^{\delta(w)\sqrt{w}} \text{ (or } \beta_2(w) \text{ below)},$$
*where $\delta(w)$ is a constant determined by $w$, as in Proposition A.2.*

PROOF. Since $\cap_i L_i^{z_i} = \emptyset$, for any $x_k \in L_k^{z_k}$, there exists some $L_j^z$ s.t. $x_k \notin L_j^{z_j}$; now, for this small group $L_j^{z_j}$, if for any $x_j \in L_j^{z_j}$ we have $h(x_k) \neq h(x_j)$, i.e., $x_k \notin h(L_j^{z_j})$, we say that $x_k$ is *collision-free*. If $|L_j^{z_j}| \leq \delta(w)\sqrt{w}$, from the union bound,
$$\mathbf{Pr}\left[x_k \text{ is } \textit{collision-free}\right] \geq 1 - \frac{|L_j^{z_j}|}{w} \geq 1 - \frac{\delta(w)}{\sqrt{w}}, \quad (7)$$
where $\delta(w)$ is defined in Proposition A.2. Note that $\cap_i h(L_i^{z_i}) = \emptyset$ implies that every $x_k$ in $L_k^{z_k}$ is collision-free. So, if furthermore $|L_k^{z_k}| \leq \delta(w)\sqrt{w}$, we have
$$\mathbf{Pr}\left[\cap_i h(L_i^{z_i}) = \emptyset\right] \geq \left(1 - \frac{\delta(w)}{\sqrt{w}}\right)^{\delta(w)\sqrt{w}}.$$

The derivation of (7) assumes independence of the randomized hash function $h$. If $h$ is generated from a random permutation $p$, i.e., taking the prefix of $p(x)$ as $h(x)$, then by considering *negative dependence* [14], a similar (a bit weaker) bound can be derived.

From Proposition A.2(iii), with probability at least $1 - \frac{1}{4\sqrt{w}}$, we have $|L_i^{z_i}| \leq \delta(w)\sqrt{w}$ for a group $L_i^{z_i}$. Given $|L_k^{z_k}| \leq \delta(w)\sqrt{w}$, there are at most $\min\{k, \delta(w)\sqrt{w}\}$ $L_j^{z_j}$'s involved in the analysis of (7). From the union bound, with probability at least $1 - \frac{1+\delta(w)\sqrt{w}}{4\sqrt{w}}$, we have $|L_j^{z_j}| \leq \delta(w)\sqrt{w}$ for all of these $1 + \delta(w)\sqrt{w}$ groups. So, with probability at least
$$\beta_1(w) = \left(1 - \frac{1+\delta(w)\sqrt{w}}{4\sqrt{w}}\right) \cdot \left(1 - \frac{\delta(w)}{\sqrt{w}}\right)^{\delta(w)\sqrt{w}},$$
(notice the independence between $g$ and $h$) we have $\cap_i h(L_i^{z_i}) = \emptyset$.

If we use Proposition A.2(ii) to bound the probability of $|L_k^{z_k}| \leq 3\sqrt{w}/2$ (then there are at most $\min\{k, 3\sqrt{w}/2\}$ $L_j^{z_j}$'s involved in the analysis of (7)), we can derive a tighter bound in a similar way:
$$\beta_2(w) = \left(1 - \exp\left(-\frac{\sqrt{w}}{12}\right) - \frac{3}{8}\right) \cdot \left(1 - \frac{\delta(w)}{\sqrt{w}}\right)^{3\sqrt{w}/2}. \quad \square$$

Thus, we have shown the probability of *successful filtering* ($\beta_1(w)$ or $\beta_2(w)$ as a conservative lower bound) is at least a constant depending only on the machine-word width $w$ (but independent on the number and the sizes of sets), and increases with $w$. It can be magnified to $1 - (1 - \beta(w))^m$ by using $m > 1$ word images of independent hash functions for filtering.

### A.5.2 *Filtering Performance in Practice*

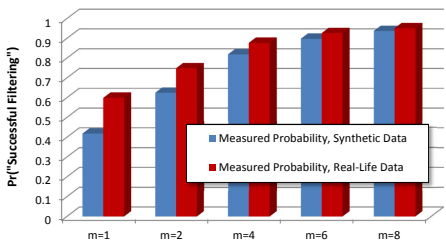

**Figure 9: Filtering Performance in Experiments**

In this section we evaluate the efficiency of the word images for filtering. In Figure 9, we have plotted the probability that, for different numbers $m$ of hash functions, a given pair of small groups with an empty intersection is filtered; as before, we use $w = 64$. As the datasets, we use the synthetic data from the first experiment in Section 4 (with an intersection size of 1% of the set size) and the 2-word queries described in the experiments on real data derived from Bing/Wikipedia. As we can see, the probabilities are very similar for both datasets, with slightly better filtering performance for the asymmetric real data. Moreover, the real-life successful-filtering probabilities are significantly better than the theoretical bounds derived in Lemma A.1 and Lemma A.3 (where $m = 1$).

### A.5.3 *Proof of Theorem 3.9*

For any $z_k \in \{0,1\}^{t_k}$, let $z_i$ be its $t_i$-prefix (as in Algorithm 5). For computing $\cap_i L_i^{z_i}$, there are two cases:

(i) If $\cap_i L_i^{z_i} = \emptyset$, from Lemma A.3, we have $\cap_i h_j(L_i^{z_i}) \neq \emptyset$ with probability at most $1 - \beta(w)$, and thus $\cap_i h_j(L_i^{z_i}) \neq \emptyset$ for all $j = 1, \ldots, m$ with probability at most $(1 - \beta(w))^m$. So we *wastefully* compute $\cap_i L_i^{z_i}$ with probability at most $(1 - \beta(w))^m$.

(ii) If $\cap_i L_i^{z_i} \neq \emptyset$, we must have $\cap_i h_j(L_i^{z_i}) \neq \emptyset$ for all $j$, and compute $\cap_i L_i^{z_i}$. Case (ii) happens at most $r = |\cap_i L_i|$ times.

We compute $\cap_i L_i^{z_i}$ using the linear merge algorithm in linear time $\mathbf{O}(\sum_i |L_i^{z_i}|)$, or $\mathbf{O}(k\sqrt{w})$ time in expectation. In case (i), since there are $n_k/\sqrt{w}$ groups in $L_k$, for all groups, this contributes a factor $\mathbf{O}(\max(n, kn_k)(1 - \beta(w))^m)$; and in case (ii), this contributes a factor $\mathbf{O}(kr\sqrt{w})$ (since (ii) happens at most $r$ times).

We also need to test whether $\cap_{i=1}^k h_j(L_i^{z_i}) \neq \emptyset$ for all $j = 1, \ldots, m$. Since there are $n/\sqrt{w}$ groups $L_i^{z_i}$'s, with careful memorization of partial results (e.g., reusing $h_j(L_1^1) \cap h_j(L_2^{10})$ when computing $h_j(L_1^1) \cap h_j(L_2^{10}) \cap h_j(L_3^{100})$ and $h_j(L_1^1) \cap h_j(L_2^{10}) \cap h_j(L_3^{101})$), this contributes a factor $\mathbf{O}(mn/\sqrt{w})$ in total.

So from the above analysis, Algorithm 5 needs a total of
$$\mathbf{O}\left(\frac{\max(n, kn_k)}{\alpha(w)^m} + \frac{mn}{\sqrt{w}} + kr\sqrt{w}\right) \quad (8)$$
time in expectation, where $\alpha(w) = 1/(1 - \beta(w))$. $\square$

## A.6 Analysis of Algorithm HashBin (Proof of Theorem 3.11)

For HashBin, the intuition is: in the resulting partitioning, we have $\mathbf{O}(1)$ element in each group $L_1^z$, and $\mathbf{O}(n_2/n_1)$ elements in each group $L_2^z$. The expected running time is:
$$\mathbf{E}\left[\sum_{z \in \{0,1\}^t} (|L_1^z| \log |L_2^z|)\right]$$
$$= \sum_{z \in \{0,1\}^t} (\mathbf{E}\left[|L_1^z|\right] \log |L_2^z|) \quad \text{(suppose } |L_2^z|\text{'s are fixed)}$$
$$= \sum_{z \in \{0,1\}^t} \log |L_2^z| \quad \text{(because } \mathbf{E}\left[|L_1^z|\right] = 1\text{)}$$
$$\leq n_1 \log(n_2/n_1) \quad (\sum_{z \in \{0,1\}^t} |L_2^z| = n_2 \text{ and } \log(\cdot) \text{ is concave}). \quad \square$$

### A.6.1 HashBin *using the Multi-resolution Structure*

Algorithm HashBin works on a simplified version of the multi-resolution data structure (Figure 2) introduced in Section 3.2.1.

Here, we use a random permutation $g : \Sigma \to \Sigma$ to partition sets into small groups. To pre-process $L_i$, we first order all the elements in $L_i$ according to $g(x)$. Then any small group $L_i^z = \{x \mid x \in L_i \text{ and } g_t(x) = z\}$ (for any $t$) corresponds to a consecutive interval in $L_i$. For each small group $L_i^z$, we only need to store its starting position $\mathrm{left}(L_i^z)$ and ending position $\mathrm{right}(L_i^z)$.

For each $x \in L_1^z$, we need to check whether $x \in L_2^z$. Suppose $L_2^z = \{x^1, x^2, \ldots, x^s\}$. Although elements in $L_2^z$ are not sorted in their own order, they are ordered as $g(x^1) \leq g(x^2) \leq \ldots \leq g(x^s)$ in the preprocessing. So to check whether $x \in L_2^z$, we can binary-search whether $g(x)$ is in $\{g(x^1), g(x^2), \ldots, g(x^s)\}$, since the random permutation $g$ is a one-to-one mapping from $\Sigma$ to $\Sigma$.

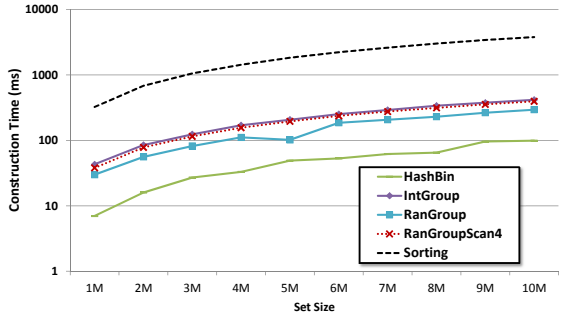

**Figure 10: Preprocessing Overhead**

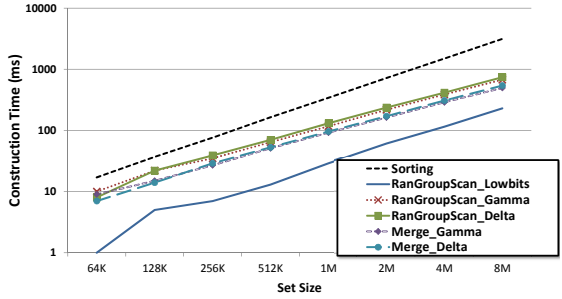

**Figure 11: Preprocessing Overhead (with compression)**

## B. COMPRESSION FOR ALGORITHM 5

For each small group $L_i^z$, we can use the standard techniques based on $\gamma$- and $\delta$-coding (see [23], p.116) to compress the elements stored sequentially at the end of the block associated with $L_i^z$. However, the decoding of $\gamma$- and $\delta$-coding is expensive.

As an alternative, we describe simple but effective (i.e., efficient in decoding) compression technique for Algorithm 5 in the following:

(i) Instead of storing the length len of each structure, we can store the size $|L_i^z|$, since the structure length len can be derived from $|L_i^z|$. As proved in Proposition A.2, $|L_i^z|$ is usually very small, so we store it using *unary code* (e.g., $011 = 2$).

(ii) Only if $|L_i^z| > 0$, we store $h_1(L_i^z), h_2(L_i^z), \ldots, h_m(L_i^z)$ in the following $m$ words.

(iii) To store elements of $L_i^z$ in the remaining part of this block, we can use the standard techniques based on $\gamma$- or $\delta$-coding. We present another compression technique here, which is specifically designed for our algorithm. Its decoding is much more efficient than $\gamma$- or $\delta$-coding. First, for the purpose of partitioning sets into small groups, we use a random permutation as $g$. Then, assuming $|\Sigma| = 2^w$ (the worst case), instead of storing each $x \in L_i^z$, we store $\mathsf{lowbits}_{t_i}(x) = g(x) \bmod 2^{w-t_i}$, i.e., the lowest $w - t_i$ bits of $g(x)$; and the remaining highest $t_i$ bits of $g(x)$ correspond to $z = g_{t_i}(x)$. In online-processing, decoding in this compression scheme can be done efficiently: to get $g(x)$ for an element $x \in L_i^z$, we concatenate $z = g_{t_i}(x)$ to $\mathsf{lowbits}_{t_i}(x)$. Since $g$ is a one-to-one mapping from $\Sigma$ to $\Sigma$, the intersection of $L_1$ and $L_2$ is equivalent to the intersection of $g(L_1)$ and $g(L_2)$.

Following is some basic analysis to establish an upper bound of the space consumed by our compression technique. Recall $t_i = \lceil \log(n_i/\sqrt{w}) \rceil$. There are $n_i/\sqrt{w}$ small groups in $L_i$ in total. Storing all of them requires:

(i) $n_i + n_i/\sqrt{w}$ bits for $|L_i^z|$'s (since $\sum_z |L_i^z| = n_i$);
(ii) at most $mw \cdot n_i/\sqrt{w}$ bits for $h_j(L_i^z)$'s; and
(iii) $(w - t_i) \cdot n_i$ bits for all elements (we store $g(x) \bmod 2^{w-t_i}$).

## C. ADDITIONAL EXPERIMENTS

### C.1 Preprocessing Overhead

In this section, we evaluate the time taken to construct the novel structures when given a set $L_i$ as input. Our approach is similar to inverted indexes (and nearly all of the competing algorithms) in that the elements have to be sorted during pre-processing; thus, to put the construction overhead in perspective, we also measure and plot the overhead of sorting using an in-memory *quicksort* (averaging the time over 10 random instances). Figure 10 shows the results for the construction time for the data structures without compression for different set sizes $|L_i|$. Note that we use a log-scale on the $y$-axis to better separate the different graphs. As we can see, the additional construction overhead is generally a small fraction of the sorting overhead.

Figure 11 shows the overhead for constructing different compressed structures. We also plot the overhead for compressing the sets without additional hash images (resulting in the structures used in the compressed Merge, i.e., Merge_Gamma and Merge_Delta). Again, the required overhead is only a small fraction of the sorting overhead; also, the preprocessing time for the Lowbits compression scheme which yields the best intersection performance in Section 4.1 is significantly lower than the alternatives.

### C.2 More Experiments on Real Data

In this section, we present a breakdown of the experiments on real data in Section 4; to understand how the number of keywords in a query affect the relative performance in this scenario, we plotted the distribution of average intersection times for 2-, 3- and 4-keyword queries separately in Figure 12. As we can see, the relative performances are similar as seen earlier with three exceptions: (a) the Merge algorithm performs worse with increasing number of keywords (as it cannot leverage the asymmetry in any way), (b) in contrast, Hash performs increasingly better, but still remains (close to) the worst performer, and (c) for 4-keyword queries, RanGroup slightly outperforms RanGroupScan.

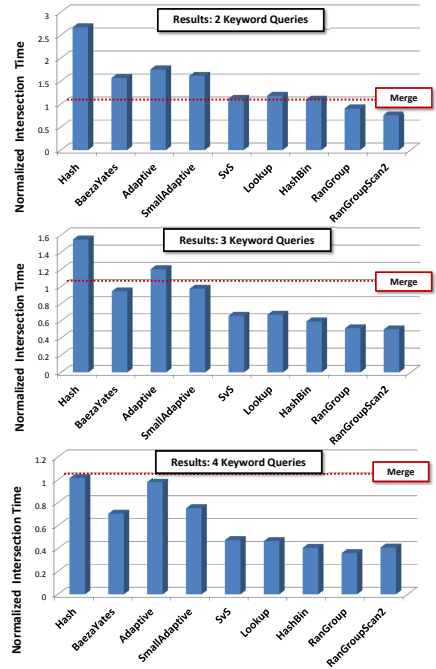

**Figure 12: Normalized Execution Time on a Real Workload**